\documentclass[twocolumn]{aastex63}
\usepackage{rotating}
\PassOptionsToPackage{colorlinks,linkcolor={blue},citecolor={blue},urlcolor={red}}{hyperref}
\usepackage{hyperref}
\usepackage{natbib}
\usepackage{float}
\usepackage{color}
\usepackage{graphicx}
\usepackage{amssymb}
\usepackage{amsmath}
\usepackage{enumitem}
\usepackage{url}
\usepackage{multirow,helvet}
\usepackage{todonotes}
\usepackage{lineno}
\usepackage{subfigure}

\revised{\today}


\def\gtrsim{\mathrel{\hbox{\rlap{\hbox{\lower4pt\hbox{$\sim$}}}\hbox{$>$}}}}
\def\lesssim{\mathrel{\hbox{\rlap{\hbox{\lower4pt\hbox{$\sim$}}}\hbox{$<$}}}}

\def\aap{A\&A}

\def\apjl{ApJL}
\def\apjs{ApJS}

\shorttitle{Radio and X-ray Observations of SNR~B0453--685 in the LMC}

\shortauthors{Eagle et al.}

\begin{document}

\title{Parkes Radio and NuSTAR X-ray Observations of the Composite Supernova Remnant B0453--685 in the Large Magellanic Cloud}

\author[0000-0001-9633-3165]{Jordan Eagle}
\affiliation{NASA Goddard Space Flight Center, Greenbelt, Maryland, 20771, USA}

\author[0000-0002-8548-482X]{Jeremy Hare}
\affiliation{NASA Goddard Space Flight Center, Greenbelt, Maryland, 20771, USA}
\affiliation{Center for Research and Exploration in Space Science and Technology, NASA/GSFC, Greenbelt, Maryland 20771, USA}
\affiliation{The Catholic University of America, 620 Michigan Ave., N.E. Washington, DC 20064, USA}

\author[0000-0002-6584-1703]{Elizabeth Hays}
\affiliation{NASA Goddard Space Flight Center, Greenbelt, Maryland, 20771, USA}

\author[0000-0002-0394-3173]{Daniel Castro}
\affiliation{Harvard-Smithsonian Center for Astrophysics,
Cambridge, MA 02138, USA}

\author[0000-0003-4679-1058]{Joseph Gelfand}
\affiliation{New York University Abu Dhabi, P.O. Box 129188, Abu Dhabi, United Arab Emirates}

\author[0000-0002-4644-6580]{Jwaher Alnaqbi}
\affiliation{New York University Abu Dhabi, P.O. Box 129188, Abu Dhabi, United Arab Emirates}

\author[0000-0002-0893-4073]{Matthew Kerr}
\affiliation{Space Science Division, Naval Research Laboratory, Washington, DC 20375-5352, USA}

\author[0000-0002-9618-2499]{Shi Dai}
\affiliation{Western Sydney University, Locked Bag 1797, Penrith South DC, NSW 2751, Australia}

\author[0000-0002-8784-2977]{Jean Ballet}
\affiliation{Universit\'{e} Paris-Saclay, Universit\'{e} Paris Cit\'{e}, CEA, CNRS, AIM, 91191 Gif-sur-Yvette, France}

\author[0000-0002-6606-2816]{Fabio Acero}
\affiliation{Universit\'{e} Paris-Saclay, Universit\'{e} Paris Cit\'{e}, CEA, CNRS, AIM, 91191 Gif-sur-Yvette, France}

\author[0000-0002-6986-6756]{Patrick Slane}
\affiliation{Harvard-Smithsonian Center for Astrophysics,
Cambridge, MA 02138, USA}

\author[0000-0002-6584-1703]{Marco Ajello}
\affiliation{Clemson University,
Clemson, SC 29634, USA}

\begin{abstract}
Gamma-ray emission is observed coincident in position to the evolved, composite supernova remnant (SNR) B0453--685. Prior multi-wavelength investigations of the region indicate that the pulsar wind nebula (PWN) within the SNR is the most likely origin for the observed gamma-rays, with a possible pulsar contribution that becomes significant at energies below $E \sim 5\,$GeV. Constraints on the PWN hard X-ray spectrum are important for the most accurate broadband representation of PWN emission and determining the presence of a gamma-ray pulsar component. The results of Parkes radio and NuSTAR X-ray observations are presented on PWN B0453--685. 
We perform a search for the central pulsar in the new Parkes radio data, finding an upper limit of $12\,\mu$Jy. A pulsation search in the new NuSTAR observation additionally provides a 3$\sigma$ upper-limit on the hard X-ray pulsed fraction of 56\%. The PWN is best characterized with a photon index $\Gamma_X = 1.91 \pm 0.20$ in the 3--78\,keV NuSTAR data and the results are incorporated into existing broadband models. Lastly, we characterize a serendipitous source detected by Chandra and NuSTAR that is considered a new high mass X-ray binary candidate.

\end{abstract}

\section{Introduction}

Pulsar wind nebulae (PWNe) are produced when the highly magnetized, relativistic particle winds formed by the conversion of rotational energy from energetic pulsars, interact with their surroundings. The evolution of PWNe depends heavily on the evolution of the central pulsar, host supernova remnant (SNR), and the structure of the surrounding interstellar medium \citep[ISM; e.g.,][]{gaensler2006}. An increasing number of PWNe are being identified at very high energies (VHE, $E>100\,$GeV), apparent in the latest TeV surveys \citep[e.g.,][]{tevcat2008} and the recent Large High Altitude Air Shower Observatory (LHAASO) catalog \citep{lhaaso2023}. The VHE emission observed from PWNe is attributed to the Inverse Compton (IC) scattering of relativistic particles off ambient photon fields such as the cosmic microwave background (CMB). The relativistic particles also radiate via synchrotron emission in the PWN magnetic field. The relativistic particle population will eventually escape into the ISM and may be contributing to the observed cosmic ray (CR) electron--positron population \citep[e.g.,][]{crflux}.

A small number ($\sim 11$) of PWNe are identified by the Fermi--LAT in the lower-energy gamma-ray band ($E < 100\,$GeV), but many currently unidentified Fermi PWN counterparts are likely entangled within the bright diffuse Galactic background and require 
a detailed analysis of each source region in Fermi--LAT data \citep[e.g.,][]{acero2013}. Such an analysis was performed on a new gamma-ray source found coincident to the evolved ($\tau \sim 14\,$kyr) composite SNR~B0453--685 located in the Large Magellanic Cloud \citep[LMC,][]{eagle2023}. A broadband investigation combining the observed gamma-ray emission to available radio and X-ray data concluded the PWN with a possible pulsar component below $E \sim 5\,$GeV is the most likely origin based on a radiative evolutionary model that accounts for the basic energy losses of the PWN as it evolves. The central pulsar has remained undetected despite comprehensive radio pulsar surveys performed on the LMC \citep[e.g.,][]{manchester2006}.

In order to determine the potential for PWNe to efficiently accelerate particles, we must be able to accurately constrain the synchrotron cut-off and MeV--GeV gamma-ray shapes. In particular, the particle cut-off energy and maximum energy as well as the properties of the ambient photon fields determine these shapes. For B0453--685, the prior report by \citet{eagle2023} uses different values for these parameters. Despite the variations in the model techniques and predicted properties, the models predict a cut-off in the synchrotron emission from B0453-685 just beyond the {\it Chandra} 0.5--7\,keV X-ray energy range. However, the {\it Chandra} X-ray spectrum for the PWN is measured to be very hard ($\Gamma_X \sim 1.7$) and does not indicate a cut-off. 

In Section~\ref{sec:radio_obs_and_dat}, we describe the new Parkes radio data reduction and analysis results in search for the pulsar. In Section~\ref{sec:obs_and_dat}, we present new 3--78\,keV NuSTAR data analysis method and results to constrain the presence of a synchrotron cut-off for the PWN B0453--685. In Section~\ref{sec:discuss} we combine the NuSTAR results to 
existing broadband representations. We discuss the implications of the models and the broadband emission and conclude in Sections~\ref{sec:discuss} and \ref{sec:conclude}. The archival {\it Chandra} observation of B0453--685 also detects an unknown second source, CXO~J045359.7--682804, faint in the hard X-ray band and located to the north of B0453--685. The new NuSTAR observations also detect faint hard X-ray emission associated to this source. The X-ray properties such as a hard spectral index imply that the source is a high-mass binary candidate. The serendipitous detection and characterization is presented here, considering X-ray (Section~\ref{sec:xray_src2}) and optical (Section~\ref{sec:opt_src2}) observations. 

\section{Parkes 64\,m Radio Observations and Data Reduction}
\label{sec:radio_obs_and_dat}


Using the ultrawideband receiver \citep{Hobbs20} on the Murriyang telescope of the Parkes Observatory, we observed PWN~B0453$-$685 on 30 Dec 2020 for two epochs (12,048\,s and 2,611\,s) and on 1 Jan 2021 for one epoch (21,434\,s) under project P1087.  The observing band range covers 704--4032\,MHz, and every 256\,$\mu$s we recorded the total intensity summed from the two input polarizations filtered into 1664 spectral channels, each 2\,MHz wide.  This format is commonly known as ``search mode''.

We reduced the data using PRESTO \citep{Ransom11}.  Specifically, we used the \texttt{rfifind} task to partially mitigate the copious radio frequency interference (RFI) present in the data.  We subsequently dedispersed the data into a series of subbands using \texttt{DDplan.py} to select the appropriate resolution in dispersion measure (DM) and time resolution (downsampling) to maintain optimal sensitivity to dispersed signals.  Although pulsars found at similar positions typically have DMs $<$100\,pc\,cm$^{-3}$, we considered DMs up to 1000\,pc\,cm$^{-3}$.  We then performed a search for periodic signals with the \texttt{accelsearch} task, which combines power in adjacent Fourier bins to account for frequency shifts of the target signal over the observation.  We did this twice, first using \texttt{zmax}=0, i.e. no acceleration, which would be appropriate for an isolated pulsar powering a pulsar wind nebula.  We also used \texttt{zmax}=200, a search which maintains sensitivity to a wide range of binary pulsars.  Because pulsars generally have soft spectra and are not expected to be as detectable above 2\,GHz as below; and due to the presence of RFI throughout the band; we also carried out a search on a subset of the data, specifically 896 MHz of bandwidth selected from 1216--2112\,MHz, a range which is relatively less contaminated by RFI.

We selected all candidates above an estimated 4$\sigma$ significance and prepared pulse profiles using \texttt{prepfold}.  We inspected each of the several hundred candidates resulting from each epoch and identified no new pulsars.  Consequently, we estimate an upper limit on the flux density assuming an ideal radiometer,
\begin{equation}
S_{\mathrm{min}} = \mathrm{(S/N)_{min}} T_{\mathrm{sys}} G \left[\epsilon\,n_p\,t_{\mathrm{obs}}\,\Delta f\,\left(W/P-1\right)\right]^{\frac{1}{2}}\mathrm{\,mJy},
\end{equation}
where we use a minimum signal-to-noise ratio of 5, $T_{\mathrm{sys}}$=21\,K, G=1.8Jy/K, $n_p$=2, a bandwidth of 886\,MHz (our narrower band), a typical fractional pulse width $W/P=0.1$, and the observing times listed above.  We define an observing efficiency $\epsilon=0.75$ to account for data lost to RFI, band taper, etc.  We obtain upper limits of 12, 16, and 34\,$\mu$Jy for the longest to shortest epochs, respectively.  We additionally searched the highest-frequency subband, about 3--4\,GHz, which is largely RFI free and would be less susceptible to scatter-broadening, but we found no compelling candidates.  Finally, we searched for single pulses, finding none, although generally, the RFI was too severe to be sensitive to these.

Although these flux upper limits are very deep, especially compared to archival surveys, because the PWN is at a much larger distance \citep[$\sim 50\,$kpc,][]{lmc2019} than a typical pulsar, the corresponding radio luminosity limits are 30--85\,(D/50\,kpc)$^2$\,mJy\,kpc$^2$.  Only about 10\% of the known Galactic pulsars are more luminous than this \citep{Manchester05}.  Consequently, we can only rule out the presence of a bright radio pulsar.

\section{X-ray Observations and Data Reduction}
\label{sec:obs_and_dat}

\subsection{Chandra}
The X-ray emission observed from B0453--685 has been extensively studied using data from both XMM--Newton and {\it Chandra} telescopes \citep[e.g.,][]{gaensler2003, haberl2012, mcentaffer2012, eagle2023}. The Chandra observations spatially resolve the PWN from the SNR, revealing soft thermal X-ray emission outlines and fills the SNR with hard non-thermal X-ray emission concentrated in the center where the PWN is located. The thermal SNR X-rays are best characterized using two \texttt{vapec} components that have $kT \sim 0.34\,$keV and $\sim 0.16\,$keV, respectively. The central PWN can be best characterized as a simple power-law with photon index $\Gamma_X \sim 1.7$. These spectral results are reported in \citet{eagle2023}. In this work, we reproduce the PWN {\it Chandra} spectral results to combine with the new NuSTAR results.

\subsection{NuSTAR}
\begin{figure}
\centering
\includegraphics[width=1.0\linewidth]{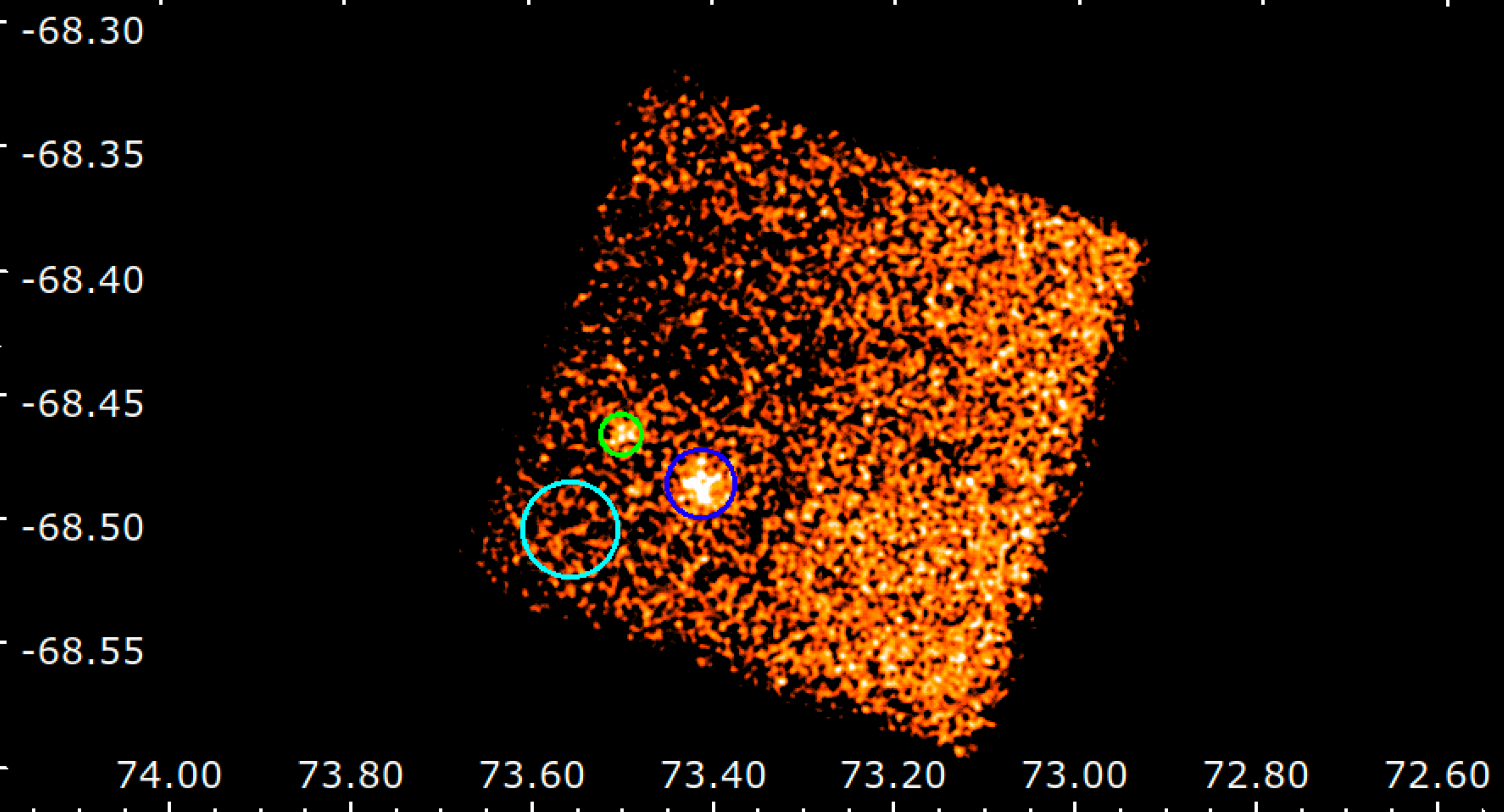}
\caption{NuSTAR FPMA detector image of B0453--685 in the 3-78 keV energy range. The image is smoothed with a Gaussian kernel.  The source (blue)  and background (cyan) regions used for spectral extraction are indicated. The source emission appears point-like, with a nearby, unrelated source ``Src2'' (green; see text for details). The coordinate grid shows the ICRS R.A. and Dec. in degrees.}\label{fig:nustar}
\end{figure}

The Nuclear Spectroscopic Array (NuSTAR; \citealt{harrison2013}) observed B0453--685 for $\approx180$\,ks on MJD 60105.99757157 (2023 10 June; ObsID: 40901001002). The data are processed and reduced using HEASOFT version 6.31 and the NuSTAR data analysis software package (NuSTARDAS) version 2.1.2 within HEASOFT \citep{heasoft}. The photon arrival times are corrected to the solar system barycenter using {\tt barycorr} and the latest NuSTAR clock correction file, which improves NuSTAR's clock accuracy to about 65 $\mu$s \citep{bachetti2021}\footnote{\texttt{nuCclock20100101v164.fits}; see \url{https://nustarsoc.caltech.edu/NuSTAR_Public/NuSTAROperationSite/clockfile.php}}. We further cleaned the data using flags {\tt saacalc=2},  {\tt saamode=optimized}, and {\tt tentacle=no}, which removes time intervals containing enhanced background emission due to the spacecraft passage through the South Atlantic Anomaly. After reducing the data, about 145\,ks of exposure are considered good time intervals. Because the PWN angular size is smaller than the NuSTAR resolution, the source spectrum and event arrival times are extracted from a 50$''$ radius circle centered on B0453--685, while background events are extracted from a 70$''$ circle, chosen to be in a source-free region on the same detector chip as the source region, see Figure~\ref{fig:nustar}. The spectra are grouped to have one count per bin and are fit using the Cash-statistic \citep{cash1979}\footnote{See \url{https://heasarc.gsfc.nasa.gov/xanadu/xspec/manual/XSappendixStatistics.html}}. The 3--78\,keV background count rate measured from FPMA is (5.6$\pm$0.3)$\times$10$^{-3}$ cts s$^{-1}$ and the background-subtracted source count rate is (3.6$\pm$0.3)$\times$10$^{-3}$ cts s$^{-1}$ measured from the regions in Figure~\ref{fig:nustar}.

\subsection{NuSTAR Spectral Analysis}\label{sec:individual}
The X-ray spectra extracted from FPMA and FPMB are fit simultaneously using the XSPEC spectral fitting package \citep{arnaud1996}. We fit the spectral data with an absorbed power-law model, using the {\tt tbabs} absorption model with {\tt wilms} abundances \citep{wilms2000}. We fix the absorbing column density to $N_{\rm H}=3.7\times10^{21}$ cm$^{-2}$ measured from {\it Chandra} 0.5--7\,keV observations \citep{eagle2023}. We additionally multiply the spectral model by a constant (where FPMA/FPMB is measured as 0.78$^{+0.09}_{-0.07}$) to account for instrumental differences between FPMA and FPMB (see e.g., \citealt{2017AJ....153....2M}). A simple power-law model anchored to FPMA provides the best-fit to the 3--78\,keV data with photon index $\Gamma_X=1.9 \pm 0.1$ for a Cash-statistic of 1444.21 for 1509 degrees of freedom. The unabsorbed flux is 4.5$^{+0.6}_{-0.5}\times10^{-13}$ erg cm$^{-2}$ s$^{-1}$ in the 3-78\,keV energy range. A broken power-law is tested to the data and yields a similar statistical fit (C-stat/dof 1440.9/1508) as the simple power-law. 
There is no evidence for the spectral cut-off predicted by the broadband models presented in \citet{eagle2023} just beyond 7\,keV.
Overall, the result is in agreement with what is found in the initial Chandra data from 0.5--7\,keV, which finds a best-fit photon index for the PWN nonthermal component of $\Gamma_X = 1.74 \pm 0.20$ and an unabsorbed flux from 0.5--7\,keV of $2.68 \pm 0.59 \times 10^{-13}$\,erg cm$^{-2}$ s$^{-1}$.   

\begingroup
\renewcommand*{\arraystretch}{1.25}
\begin{table*}
\centering
\scalebox{1.0}{
 \begin{tabular}{cccccccccc}
 \hline
 \hline
 Instruments & Energy Range & $\chi^2$/d.o.f.$^a$ & $N_H$ & $\Gamma$ & 3--8\,keV Flux\\ 
& keV & & 10$^{22}$ cm$^{-2}$ & & 10$^{-13}$ erg cm$^{-2}$ s$^{-1}$\\
\hline    
Chandra & 0.5--7 & 0.94 & 0.37$_{-0.09}^{+0.11}$ & 1.74$_{-0.20}^{+0.20}$ & 1.20$_{-0.21}^{+0.19}$\\
Chandra & 3--8  & 0.98 & 0.37 & 2.06$_{-0.34}^{+0.35}$  & 1.26$_{-0.15}^{+0.16}$ \\
NuSTAR & 3--8  & 0.99 & 0.37 & 1.93$_{-0.39}^{+0.38}$ & 1.12$_{-0.12}^{+0.14}$ \\
Chandra+NuSTAR & 3--8  & 0.93 & 0.37 & 1.96$_{-0.17}^{+0.18}$ & 1.25$_{-0.15}^{+0.16}$ \\
NuSTAR & 3--50  & 0.93 & 0.37 & 1.92$_{-0.20}^{+0.20}$  & 1.11$_{-0.11}^{+0.10}$ \\
NuSTAR & 3--78 & 0.96 & 0.37 & 1.91$_{-0.20}^{+0.20}$  & 1.20$^{+0.09}_{-0.09}$ \\
Chandra+NuSTAR & 0.5--20 & 0.99 & 0.37$_{-0.09}^{+0.08}$ & 2.01$_{-0.13}^{+0.13}$ & 1.18$_{-0.13}^{+0.15}$ \\
\hline
\hline
\end{tabular}}
\caption{Best-fit spectral model for each data set as indicated. Fits for $E> 3\,$keV are fixed to the $N_H$ value 0.37$\times10^{22}$ cm$^{-2}$. \footnotesize{$^a$measured from the Cash-statistic except for the first row, which comes from \cite{eagle2023}.}}
\label{tab:fits}
\end{table*}
\endgroup

\begin{figure}
\centering
\includegraphics[width=1.0\linewidth]{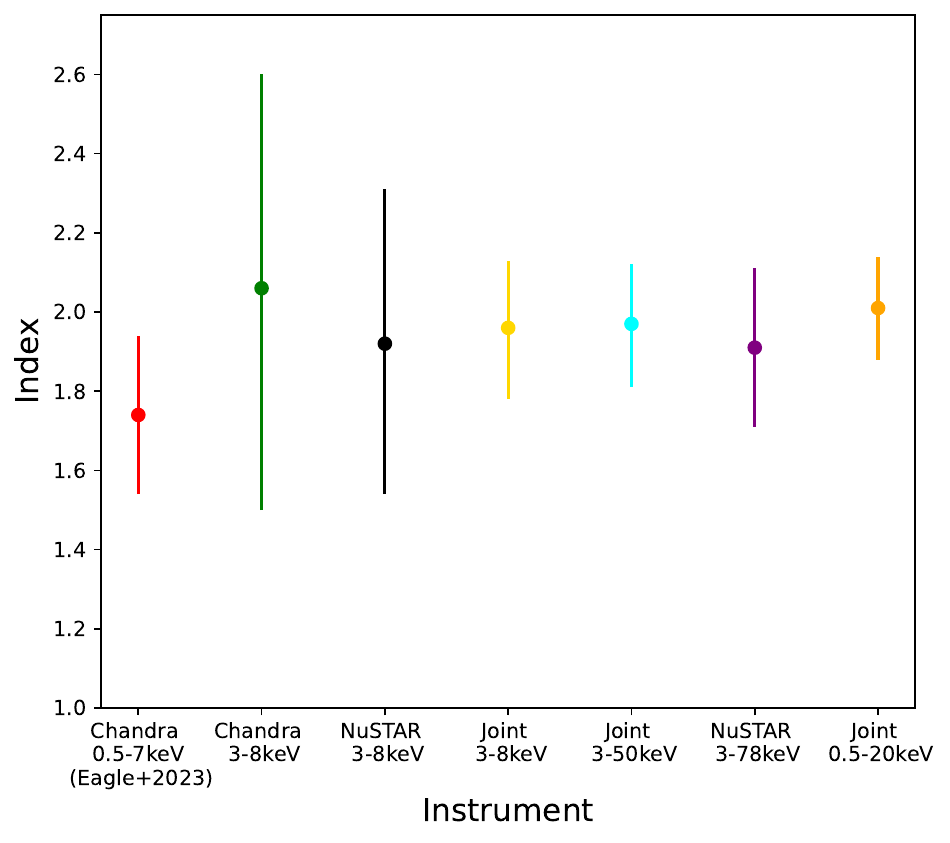}
\caption{A comparison of the power law spectral index values measured from X-ray spectral models assuming varying data sets as labeled. }\label{fig:index_values}
\end{figure}

We perform joint fits considering both the Chandra 0.5--7\,keV and NuSTAR 3--20\,keV data, after verifying that all spectral measurements are consistent and thus the cross-detector calibration differences are minimal. Figure~\ref{fig:index_values} shows the spectral measurements for varying dataset combinations and energy ranges. 
We test a combination of datasets and report the best-fit spectral values along with the 3--8\,keV unabsorbed flux values in Table~\ref{tab:fits}. The datasets are in good agreement and, within uncertainties, no significant differences in the spectral index are found, see also Figure~\ref{fig:index_values}. The NuSTAR data is not fit significantly better using a broken power-law model, finding an energy break at $16^{+3}_{-2}$\,keV, the same energy where the background begins to dominate. Similarly, the joint fits do not yield a better fit using a broken power-law over the simple power-law spectrum.  If a break in the PWN spectrum exists, it is beyond 10.4\,keV, derived by setting $\Gamma_2 = \Gamma_1 + 1$ in the 0.5--20\,keV broken power-law spectral model. 

\begin{figure*}
\begin{minipage}[b]{.5\textwidth}
\centering
\includegraphics[width=1.\linewidth]{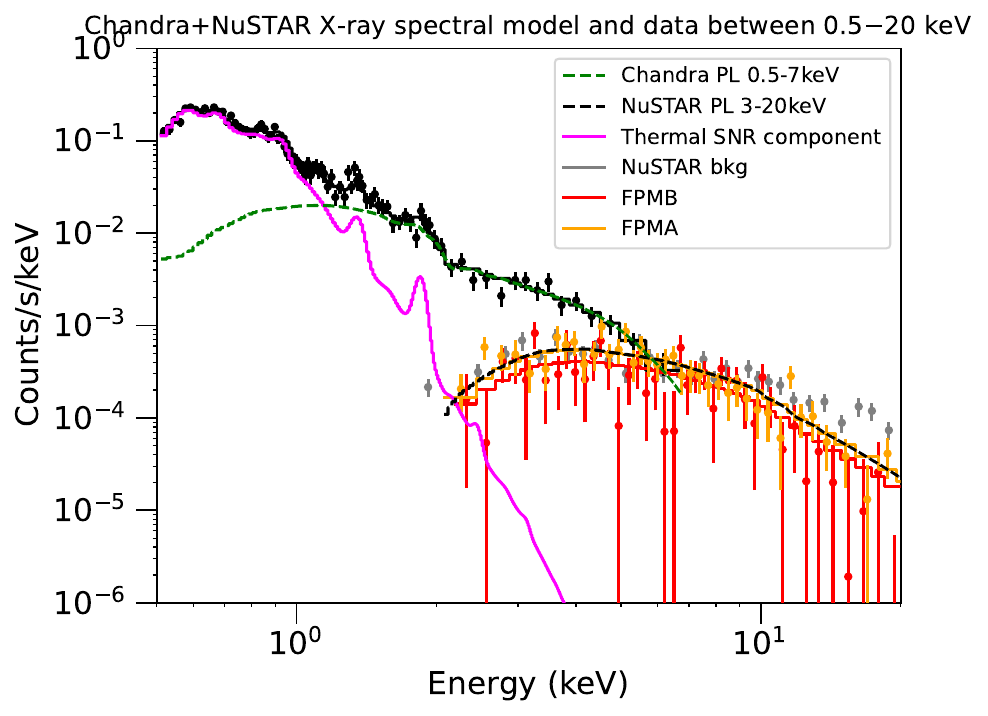}
\end{minipage}
\begin{minipage}[b]{.5\textwidth}
\centering
\includegraphics[width=1.07\linewidth]{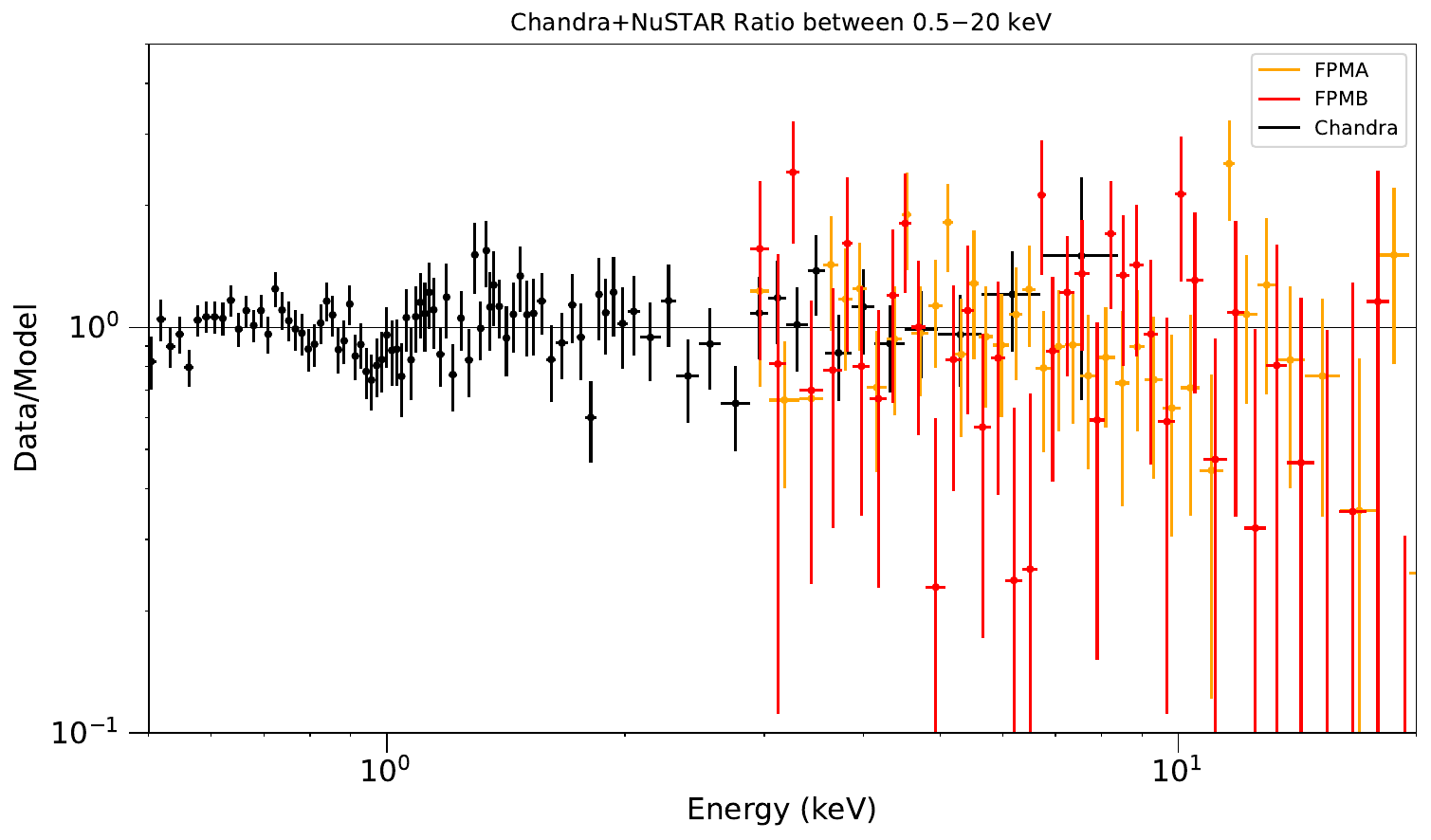}
\end{minipage}
\caption{{\it Left:} The joint fit for Chandra data from 0.5--7\,keV and the NuSTAR data from 3--20\,keV for B0453--685. The PWN 0.5--20\,keV X-ray spectrum can be fit assuming a simple power-law with a photon index $\Gamma_X = 2.01 \pm 0.13$ plus a low-energy thermal component arising in the SNR. The Chandra and NuSTAR data points are background-substracted. The background data is shown in grey for NuSTAR FPMA. The solid and dashed lines represent the source spectral models and components. {\it Right:} The residuals of the joint best-fit. In both panels, the data is grouped to 20 counts per bin for clarity.}\label{fig:joint_fit}
\end{figure*}

Given the good agreement found between the tested dataset configurations (Table~\ref{tab:fits}), we plot the 0.5--20\,keV best-fit for Chandra and NuSTAR data in Figure~\ref{fig:joint_fit}. As in \citet{eagle2023}, a thermal SNR component exists below $ E < 2$\,keV, and is therefore modeled using the same thermal model constructed and reported in their Tables~1 and 2, consisting of two \texttt{vapec} components with $kT \sim 0.34\,$keV and $\sim 0.16\,$keV, respectively. FPMA and FPMB spectra are anchored to the Chandra spectrum using the measured constants $c_1 = 0.52^{+0.09}_{-0.08}$ and $c_2 = 0.41^{+0.09}_{-0.07}$, respectively. Finally, the pulsar powering the PWN of B0453--685 has yet to be detected, therefore no additional spectral component is motivated. The SNR in X-ray is thermal in origin \citep[see also][]{gaensler2003,haberl2012,mcentaffer2012} and is negligible in the NuSTAR band such that the observed hard X-ray emission is attributed to the PWN. 

\subsection{NuSTAR Timing}
\label{sec:time}
As in Section~\ref{sec:radio_obs_and_dat}, we attempt to search for pulsations in the NuSTAR data from the central pulsar. The B0453--685 source spectrum begins to fall below the background around 10\,keV. Therefore, we limit our pulsation searches to energies between 3--10\,keV and use the $Z^{2}_m$ test, with $m$ set to 1 \citep{1983A&A...128..245B}. We search frequencies between $\nu=1.1\times10^{-5}-100$ Hz, where the lowest frequency is set by the length of the observation and the highest frequency is chosen to be 100 Hz, as young pulsars ($\tau < 100\,$kyr) typically do not have spin periods less than 10\,ms. The frequencies are over-sampled by a factor of 10, to ensure no peaks are missed. The false alarm probability can be calculated as $e^{(-Z^2_{1,max}/2)}$ multiplied by the number of trials, where $Z^2_{1,max}$ is the maximum value found in the periodogram. The number of trials is defined as $N_{\rm trial}=(\nu_{\rm max}-\nu_{\rm min})T_{\rm span}\approx1.8\times10^{7}$, where $T_{\rm span}$ is the duration of the observation (see e.g., \citealt{2021ApJ...923..249H} for additional details). This leads to a 3$\sigma$ significance threshold of $Z^2_{1,3\sigma}=45.4$.

The largest $Z^2_{1,\rm max}=35.35$, which corresponds to a $<1\sigma$ significance, was found at a frequency of 29.5679 Hz (or 33.8\,ms). Once the largest peaks are found in our chosen frequency range, we allow the high end of the energy range to vary between 5 and 15\,keV to see if the signal significance increases (due to  increasing/decreasing source/background contributions, see Figure \ref{fig:joint_fit}). We find that the significance of the 29\,Hz signal does slightly increase to $Z^2_{1,\rm max}=38.81$ when using photons with energies between 3--7\,keV. However, this signal still has a significance $<2\sigma$, therefore we conclude that no pulsations are detected from this source in the existing NuSTAR observations. 

Assuming sinusoidal pulsations, we can estimate the $3\sigma$ upper-limit on the observed pulsed fraction using $Z^2_{1,3\sigma}=N_{\text{tot}}p^2_{\text{amp}}/2+2$, where $N_{\text{tot}}$ is the total number of photons and $p_{\text{amp}}$ is the observed pulsed fraction (see e.g., \citealt{2021ApJ...923..249H}). The $3\sigma$ upper-limit on the observed pulsed fraction is 27$\%$ in the 3--10\,keV energy range. This can be converted into an intrinsic pulsed fraction by multiplying by a factor $N_{\text{tot}}/N_{\text{src}}=2.1$, where $N_{\text{src}}$ is the number of source photons (see e.g., \citealt{2023ApJ...951...80H}). We find a $3\sigma$ upper-limit on the intrinsic pulsed fraction of 56$\%$.

\begin{figure}
\centering
\includegraphics[width=1.0\linewidth]{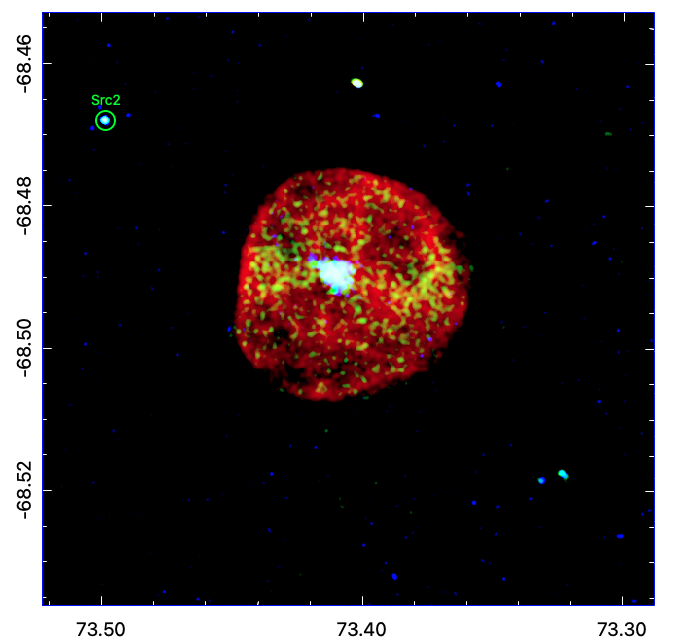}
\caption{Tri-color Chandra X-ray flux map of B0453--685. Red = 0.5--1.2\,keV, green is 1.2--2\,keV and blue is 2--8\,keV. Src2 is marked in green.}\label{fig:src2_chandra}
\end{figure}

\subsection{CXO J045359.7-682804 (Src2)}\label{sec:xray_src2}
\begin{figure*}
\begin{minipage}[b]{.5\textwidth}
\centering
\includegraphics[width=1.\linewidth]{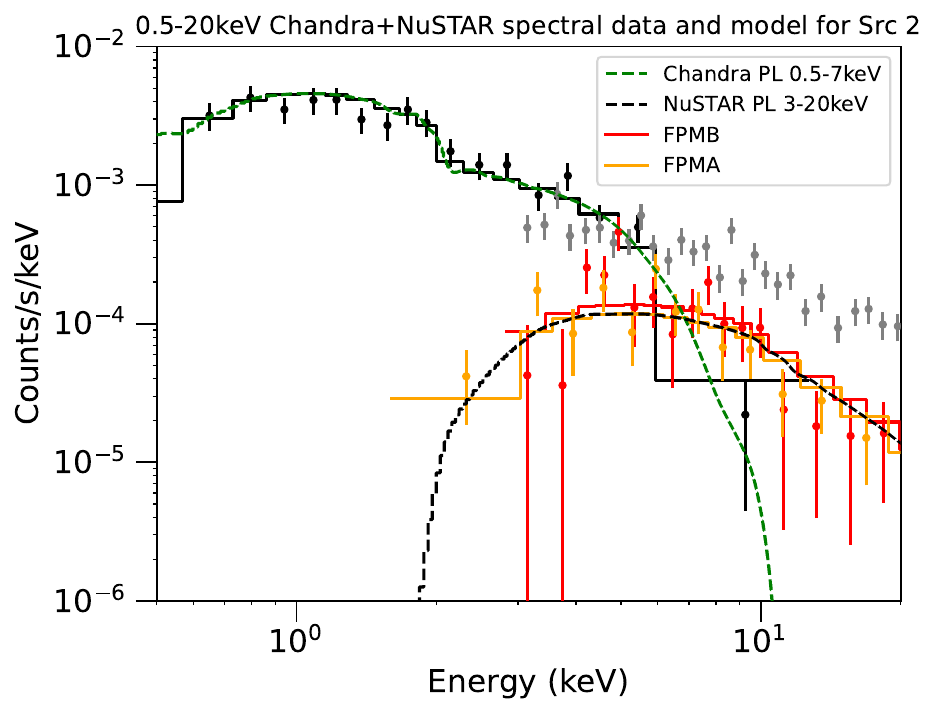}
\end{minipage}
\begin{minipage}[b]{.5\textwidth}
\centering
\includegraphics[width=1.0\linewidth]{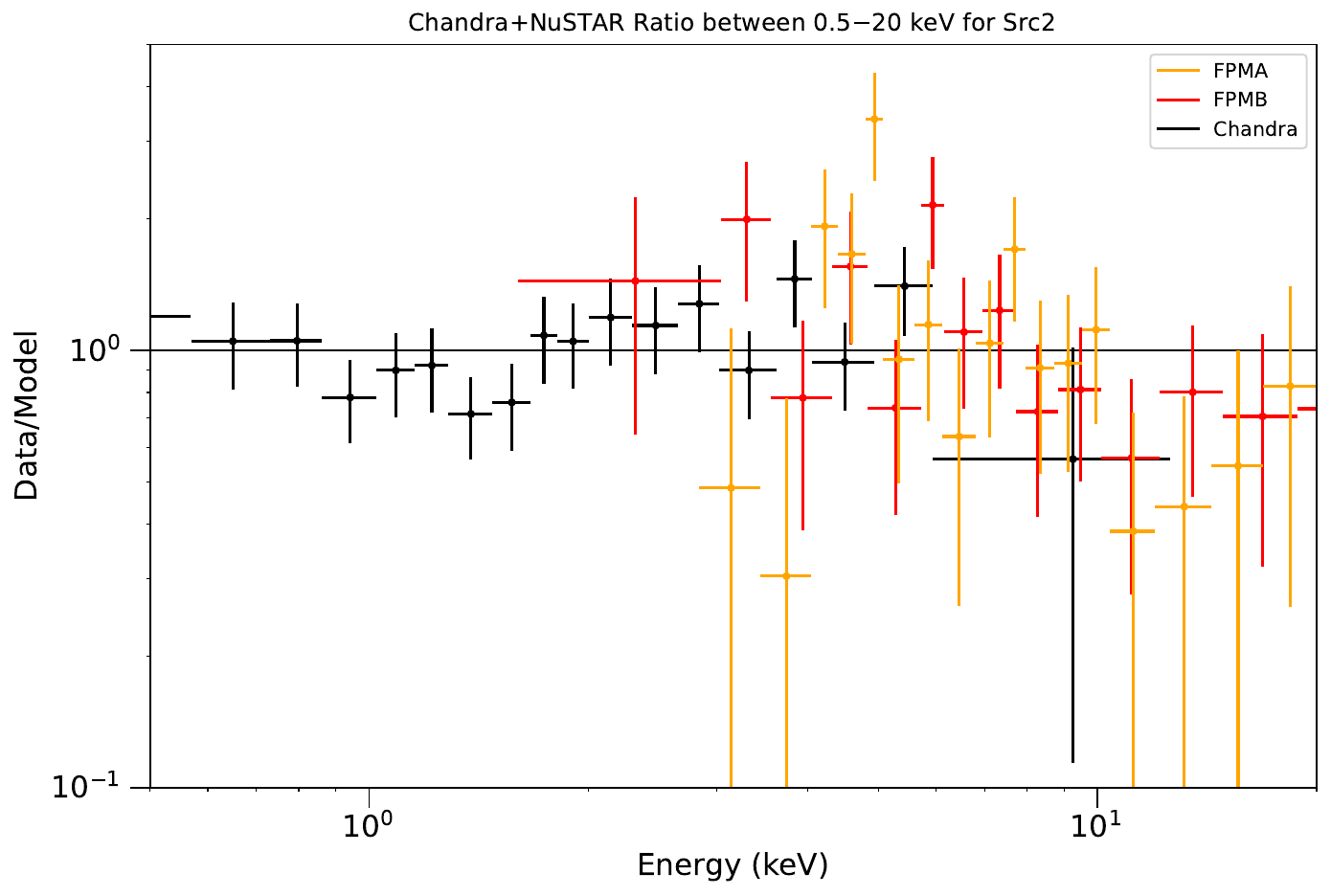}
\end{minipage}
\caption{{\it Left:} The joint fit for Chandra data from 0.5--7\,keV and the NuSTAR data from 3--20\,keV of Src2. The 0.5--20\,keV X-ray spectrum can be fit assuming a simple power-law with a photon index $\Gamma_X = 1.23 \pm 0.19$. The Chandra and NuSTAR data points are background-substracted. The background data is shown in grey for NuSTAR FPMA. The solid and dashed lines represent the source spectral models and components. {\it Right:} The residuals of the joint best-fit. In both panels, the data is grouped to 20 counts per bin for easier visualization.}\label{fig:src2_joint_fit}
\end{figure*}

A second source is detected by Chandra and NuSTAR, CXO J045359.7--682804, which we hereafter refer to as ``Src2'' (see Figure~\ref{fig:nustar}). This source appears in the Chandra data as a faint, point-like, hard X-ray source to the northeast of both the PWN and the SNR shell, see Figure~\ref{fig:src2_chandra}. The source is located at R.A., Dec equatorial coordinates (J2000) = (73.4988, --68.4679)$^\circ$ with an estimated 90\% uncertainty that is $r_{90} = 0.145^{\prime\prime}$ \citep{kim2007}. 
Chandra has a 90\% absolute astrometric uncertainty of 0.8$''$\footnote{\url{https://cxc.harvard.edu/cal/ASPECT/celmon/\#offset_history}}, which we add in quadrature to the statistical uncertainty, leading to a 90\% positional uncertainty of  0.81$''$. 
We model the Src2 spectrum in both Chandra and NuSTAR from 0.5--7\,keV and 3--78\,keV, respectively. In 0.5--7\,keV, Src2 is faint with an unabsorbed flux in the 3--8\,keV range $6.87^{+1.20}_{-1.16} \times 10^{-14}$\,erg cm$^{-2}$ s$^{-1}$. The best-fit spectrum is a simple power-law with a hard index $\Gamma_X = 0.95^{+0.24}_{-0.16}$. The $N_H$ value is considerably lower than the one measured for both B0453--685 ($N_H \sim 3.7 \times 10^{21}$\,cm$^{-2}$) and the LMC ($N_H = 2.21 \times 10^{21}$\,cm$^{-2}$), indicating it may be a foreground object, $N_H < 0.8 \times 10^{21}$\,cm$^{-2}$ (90\% confidence level). We fit the NuSTAR spectrum of Src2 with a power-law, fixing the absorption value to $0.8 \times 10^{21}$\,cm$^{-2}$. The best-fit power-law model has a photon index $\Gamma=1.7\pm0.3$, a 3--8\,keV unabsorbed flux $6.25^{+1.29}_{-1.18} \times10^{-14}$\,erg cm$^{-2}$ s$^{-1}$, and is well fit, having C-stat$=690.9$ for 738 degrees of freedom. Performing a joint fit on the Chandra and NuSTAR data as is done for B0453--685, see Figure~\ref{fig:src2_joint_fit}, we find the 0.5--20\,keV best-fit index is $\Gamma=1.23\pm0.19$, $N_H = 0.9^{+0.7}_{-0.6} \times 10^{21}$\,cm$^{-2}$, and a 3--8keV unabsorbed flux $5.68^{+1.0}_{-0.84} \times 10^{-14}$\,erg cm$^{-2}$ s$^{-1}$.  The source is not related to B0453--685, but appears to be an unidentified hard X-ray source observed by Chandra and NuSTAR. 

\section{Optical Monitoring Campaign of Src2}\label{sec:opt_src2}
The closest cataloged source in SIMBAD\footnote{\url{https://simbad.cds.unistra.fr/simbad/}} to Src2 is a star, MACHO~45.2124.10, off-set 38$^{\prime\prime}$from the X-ray position. Src2 has a more likely Gaia counterpart (GAIA DR3 4655520733857508352; \citealt{gaiamission,gaiadr3,gaiadr3paper}) offset from the X-ray source position by only 0.04$''$. This optical counterpart is also detected in NIR by 2MASS (2MASS 04535971-6828047; \citealt{2MASS}) and UV by GALEX (GALEX J045359.8-682804; \citealt{galex}). The GAIA counterpart is reported as a variable star with 100\% probability on the object type 
with an effective temperature $T_{\text{eff}} \sim 3.1 \times 10^{4}$\,K, a mean $G$ apparent magnitude $\sim 15$, and blue color, $BP-RP=-0.08$. Unfortunately, the Gaia parallax is not well measured so the distance is unknown. However, the low hydrogen column density suggests that the source may be foreground to the LMC, making it unlikely to be extragalactic. The $N_H$ also provides an estimate on the interstellar extinction $A_V = 0.41 \pm 0.27$ and corresponding $E(B-V) = 0.13 \pm 0.09$ \citep{guver2009}. The StarHorse2 catalog \citep{gaiastars2022} reports similar values for the GAIA source, $A_V = 0.55$, for a 10.9\,$M_\odot$ star at a distance $\sim 44$\,kpc.

Based on the very blue color and high effective temperature, the star is likely an O or B type star, implying that it would need to be at a minimum distance of about 10 kpc in order to have an absolute magnitude consistent with these spectral types \citep{2018A&A...616A..10G}, ignoring the low extinction and any contribution from an accretion disk. This places a lower-limit on the 3--10\,keV X-ray luminosity $L_X>10^{33}$ erg s$^{-1}$. If instead the source resides in the LMC, the 3--10\,keV X-ray luminosity would be $L_X\approx10^{34}$ erg s$^{-1}$. An optical spectrum of the source could help to confirm the spectral type of the star, and allow for a more constraining distance estimate. The optical variability, high effective temperature, hard X-ray spectral index, and estimated X-ray luminosity make Src2 a new high mass X-ray binary candidate \citep{reig2011,ferrigno2023,fortin2023}. 

\begin{figure}
    \centering
    \includegraphics[width=0.44\textwidth]{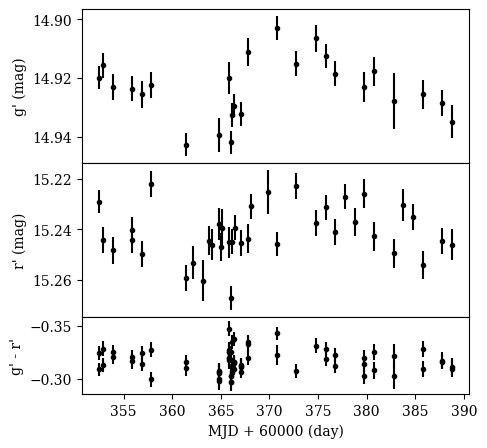}
    \caption{\textit{Upper and middle panels:} Optical light curves of Src2 in $g^\prime$ and $r^\prime$ filters between MJD 60352 and 60388 (2 December 2024 to 19 March 2024). \textit{Lower panel:} $g^\prime-r^\prime$ color evolution in the same duration.}
    \label{fig:LC}
\end{figure}

Currently, available X-ray observations do not provide enough information to detect or characterize the variability of this source. To better characterize the optical variability, we conducted a  monitoring campaign of the source using the global telescope network of Las Cumbres Observatory (LCO) in SDSS $g^\prime$ and $r^\prime$ filters between MJD 60352 and 60388 (2023 December 2 to 2024 March 19) using 30 second exposures with a 1.0-meter telescope within the network. The details of the filters used are listed in Table \ref{tab:LCO}. The data are extracted and calibrated by the X-ray Binary New Early Warning System (XB-NEWS) real-time data analysis pipeline \cite[]{Russell_2019,Goodwin_2020}. The resulting observed (reddened) light curves for both filters and the corresponding color ($g^\prime-r^\prime$) are shown in Figure~\ref{fig:LC}.

\begin{table}[h]
\centering
\begin{tabular}{cllcl} \hline\hline
\multicolumn{1}{l}{Station} & Filter & $\lambda$ (\AA) & \multicolumn{1}{l}{MJD (day)} & $\bar{m}$ (mag) \\ \hline
\multirow{2}{*}{LCO \footnote{\url{https://lco.global/observatory/}}} & $g^\prime$ & 4722 & \multirow{2}{*}{60352 - 60388} & 14.92 $\pm$ 0.01 \\
 & $r^\prime$ & 6215 &  & 15.24 $\pm$ 0.01\\\hline
\end{tabular}
\caption{Optical monitoring details of Src2. $\lambda$ is the central wavelength of a filter, and $\bar{m}$ is the average observed (reddened) magnitude.}
\label{tab:LCO}
\end{table}

\begin{figure}
    \centering
    \subfigure[]{\includegraphics[width=0.5\textwidth]{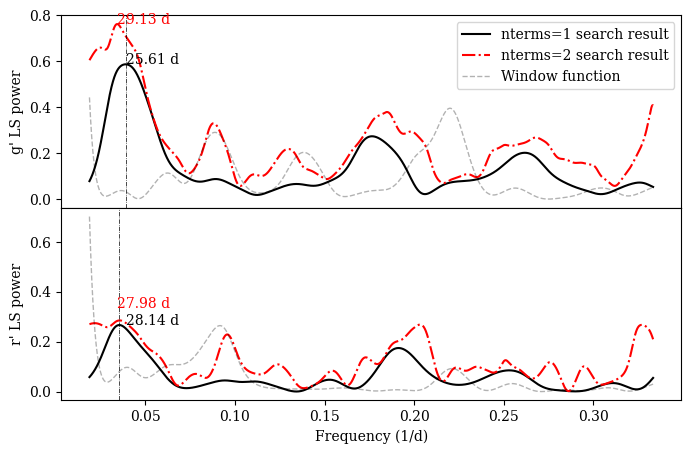}\label{fig:LS}
    }
    \subfigure[]{\includegraphics[width=0.5\textwidth]{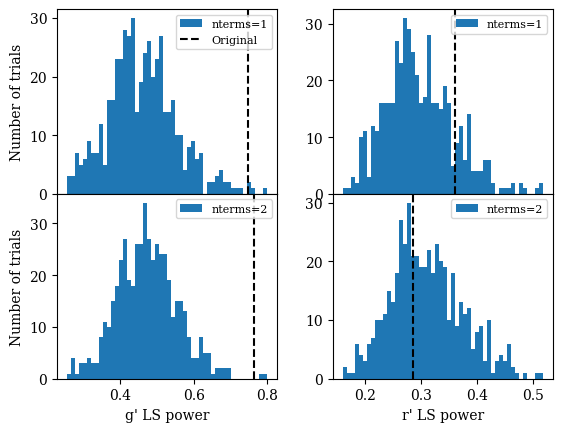}
    \label{fig:hist}
    }
    \caption{(a) LS periodograms of Src2 in $g^\prime$ and $r^\prime$ filters using nterms=1 and nterms=2 with the window function generated using nterms=1 overlaid on each periodogram. The dominant period in each is labeled. (b) Histograms of the most dominant signal power in $g^\prime$ and $r^\prime$ generated by randomizing the optical light curve of the source 500 times, using 1 or 2 nterms. The dashed vertical line selects the power resulting from the original light curve. 
    }
\end{figure}

We use the Lomb-Scargle algorithm \citep[LS,][]{Lomb_1976,Scargle_1982,VanderPlas_2018} to search for periodic variations on timescales between 3 and 53 days in the optical emission of the source. This tool contains the ``nterms'' parameter that allows adjusting the number of terms to use in the Fourier fit, and Figure \ref{fig:LS} shows the resulting periodograms. We calculate the best estimated period in a periodogram as the inverse of the frequency corresponding to the maximum LS power. We calculate periods of 29.13 and 25.61 days for $g^\prime$ and 27.98 and 28.14 days for $g^\prime$ using nterms of 1 and 2, respectively. 

To quantify the significance of each period, we first generate periodograms of the “window function'', i.e., strength of periodicity resulting from the frequency of taking measurements, using nterms=1. If the strongest peak in a filter's periodogram appears at a frequency with a comparable strength in its window function, then that peak is found to be unreliable. We overlay the window function periodograms over each filter periodogram in Figure \ref{fig:LS}. The most dominant peak in each filter does not have a corresponding significant window function signal, and hence, does not result from the cadence of these observations. The significance of the best calculated period or maximium power is given by the confidence probability (CP). CP \citep[FAP in][]{Baluev_2008} is the probability of measuring a peak of a given height (or higher) conditioned on the assumption that the data consists of Gaussian noise with no periodic component. The CP results for $g^\prime$ and $r^\prime$ suggest confidence levels of $98.39\%$ and $99.99\%$ for $g^\prime$, and $32.78\%$ and $47.39\%$ for $r^\prime$ using nterms=1 and 2, respectively. These results imply that the period calculated using $g^\prime$ data is likely real, unlike $r^\prime$.

\begin{figure}
    \centering
\includegraphics[width=1.0\linewidth]{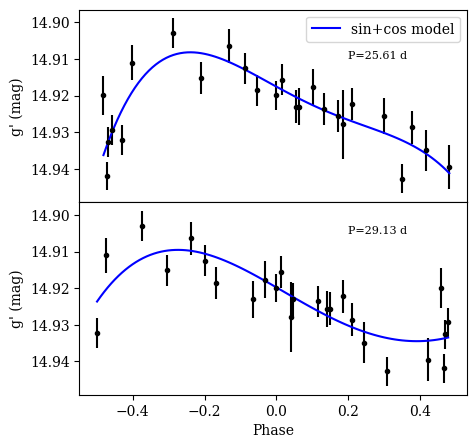}
    \caption{$g^\prime$ light curve of Src2 folded on periods calculated using $g^\prime$ data (25.61 and 29.13 days). The folding period used is labeled in each plot. Each FLC is fitted with a $\sin{}+\cos{}$ model using nterms=2.}
    \label{fig:FLC}
\end{figure}

In Figure \ref{fig:hist} we show histograms of the maximum power of a periodogram generated by randomizing the magnitudes of each filter light curve 500 times while fixing the MJD. While the figure suggests that the $r^\prime$ period has $\sim 50\%$ probability of being produced by chance, the $g^\prime$ power histogram implies a high probability ($>99\%$) for the $g^\prime$ periods to be intrinsic to the source. 
We fold the $g^\prime$ light curve on the periods we calculate using $g^\prime$ data in Figure \ref{fig:FLC}, and fit each with a $\sin{}+\cos{}$
model\footnote{\url{https://docs.astropy.org/en/stable/timeseries/lombscargle.html}}. The parameters of the model include amplitude ($mag$), offset ($phase$), and wave frequency ($phase$). The $g^\prime$ shows good agreement (reduced $\chi^{2}\sim 1.98$ and $2.2$)
with the resulting models. The periodic behavior we observe in Src2 suggests that it is likely an X-ray binary system (see e.g., \citealt[]{Zhang_2024}) with an orbital period between $\sim 25-29$\,days. 

\section{Discussion}\label{sec:discuss}
\citet{eagle2023} presents a detailed multiwavelength analysis using two different broadband modeling techniques to explore the most likely origin of the observed $\gamma$-ray emission. The Chandra X-ray data analysis finds a hard nonthermal component from the PWN and is combined with radio and the Fermi--LAT data in order to derive a broadband representation for the system. The authors find that the most plausible origin is the PWN within the middle-aged SNR~B0453--685 and possibly a substantial pulsar contribution to the low-energy $\gamma$-ray emission below $E < 5$\,GeV. Two modeling techniques are explored: one that is time-independent (NAIMA) and the other that accounts for basic energetic losses as the PWN evolves and the pulsar continuously injects particles into the PWN \citep{gelfand2009}. The latter predicts a MeV bright pulsar contribution up to $E \sim 5\,$GeV which becomes PWN-dominated above that energy. Both broadband models predict a synchrotron cut-off just beyond 7\,keV. 

\begin{figure*}
\begin{minipage}[b]{.5\textwidth}
\centering
\includegraphics[width=0.98\linewidth]{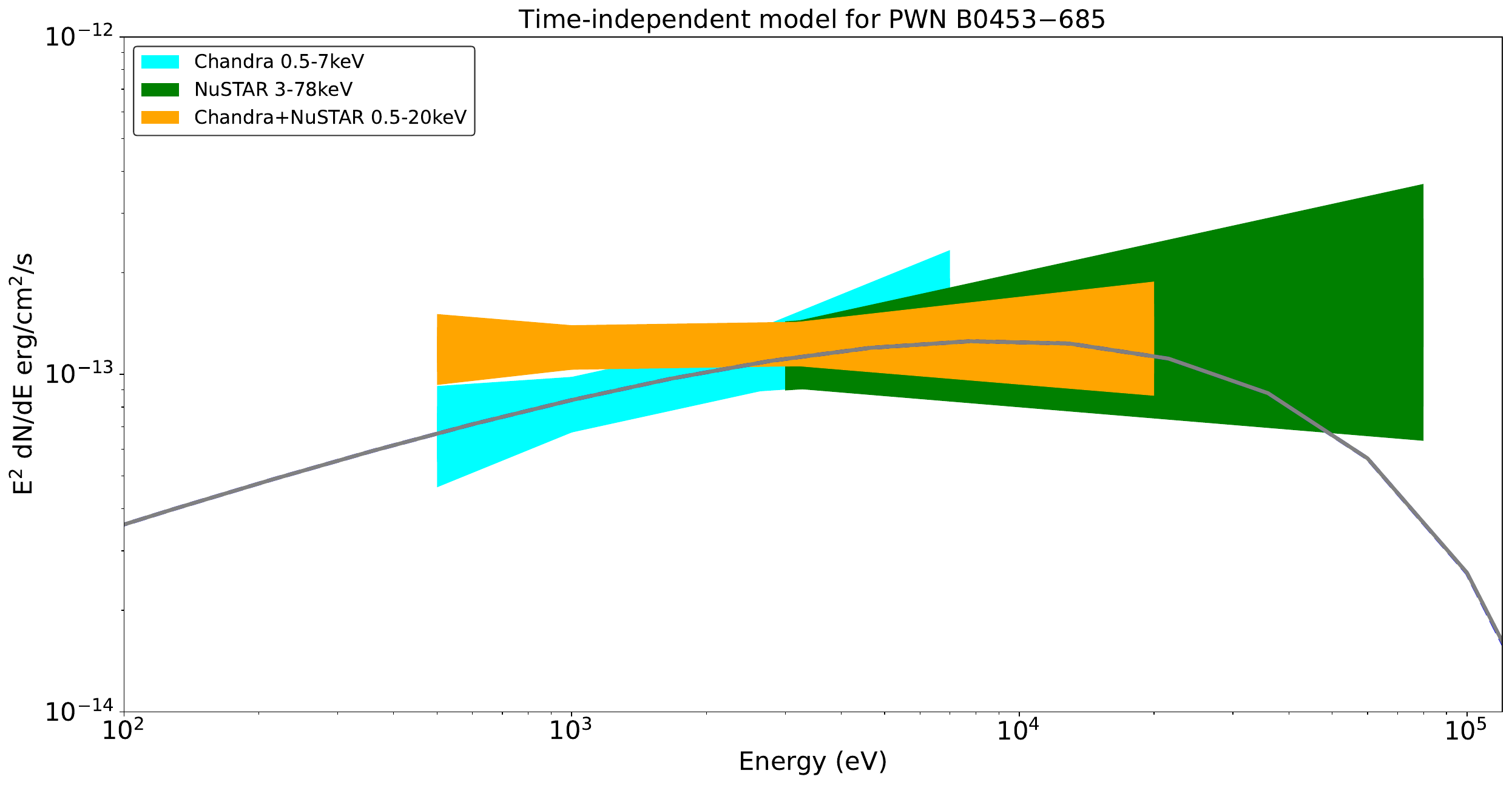}
\end{minipage}
\begin{minipage}[b]{.5\textwidth}
\centering
\includegraphics[width=1.0\linewidth]{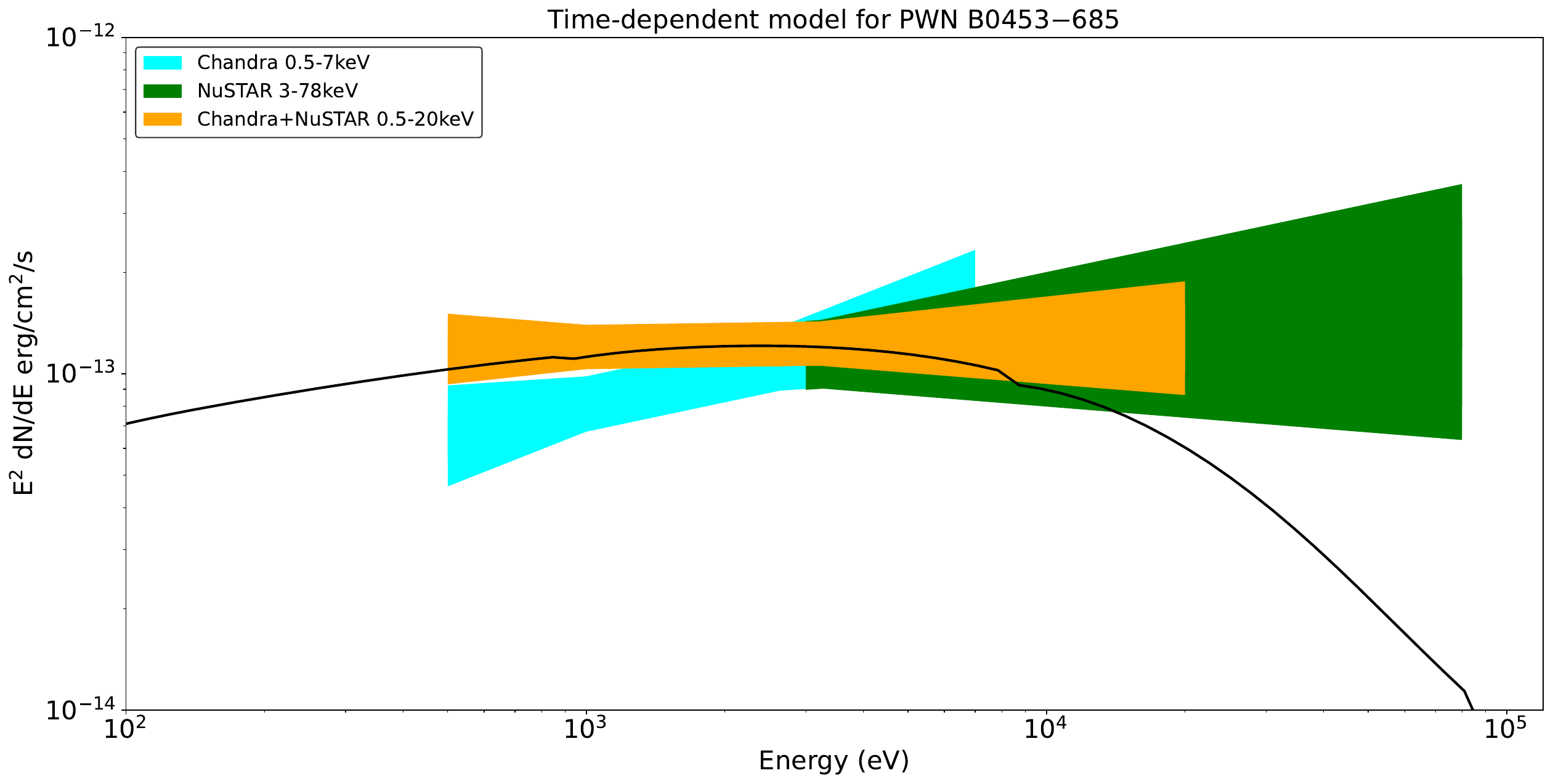}
\end{minipage}
\caption{{\it Left:} Time-independent model for B0453--685 in the X-ray band. X-ray data of the PWN (cyan) are from \citet{eagle2023} and the orange and green X-ray data are from Section~\ref{sec:individual}. 
{\it Right:} The time-dependent (evolutionary) model from \citet{eagle2023} in the X-ray band. The data is the same as in the left panel. 
}\label{fig:bb_plots}
\end{figure*}

If a synchrotron cutoff exists, the properties of the highest-energy particles can be accurately determined. The synchrotron cut-off provides the maximum particle energy \citep[e.g.,][]{temim2015} and consequently the MeV spectrum of the PWN \citep[e.g.,][]{gelfand2019}. Moreover, particle properties such as the particle energy break and the total energy output are also constrained from the synchrotron cut-off which, in turn, more accurately characterizes the shape of the particle spectrum as well as the central pulsar input. Current broadband models infer these properties by considering the radio and 0.5--7\,keV Chandra data together with the newfound Fermi--LAT $\gamma$-ray emission, alongside the predicted combination of neutron star, pulsar wind, supernova explosion, and ISM properties from the evolutionary model that agrees with the observations \citep{eagle2023}. 
The new NuSTAR results are obtained in order to characterize the synchrotron cut-off of the PWN predicted by both modeling techniques just beyond 7\,keV, but no cut-off is significantly detected. The NuSTAR data is overlaid with the existing broadband models of the PWN from \citet{eagle2023} in both panels of Figure~\ref{fig:bb_plots}. The NuSTAR data is in reasonable agreement with both model predictions, despite not characterizing a synchroton cut-off. Notably, however, the time-dependent model (right panel of Figure~\ref{fig:bb_plots}) does not describe the new X-ray results as well as the time-independent model (left panel of Figure~\ref{fig:bb_plots}). The lack of a synchrotron cut-off may be able to rule out the time-dependent model prediction.


\section{Conclusion}\label{sec:conclude}
New Parkes radio and NuSTAR hard X-ray observational results are reported in search for the central pulsar and a synchrotron cut-off in the spectrum of the PWN within B0453--685.  No bright radio pulsar is identified. The NuSTAR X-ray emission is found to be nonthermal in origin and point-like. While no synchrotron cut-off is significantly detected with NuSTAR, it does not rule out the existence of one. The background in the NuSTAR data dominates above $\sim 16\,$keV and it is possible the PWN spectral cut-off occurs beyond this energy. There is a slight hint of spectral softening both in the NuSTAR data and in joint Chandra and NuSTAR spectral fits, but within uncertainties, there is no compelling spectral break evident. The lack of a clear spectral break for an evolved PWN such as B0453--685 is surprising. The SNR in the X-ray exhibits both a thermal shell and filled center, indicating the reverse shock has interacted with the PWN already. The PWN compression from the passage of the reverse shock increases the magnetic field strength, resulting in the shortened cooling time of the high-energy (X-ray emitting) particles \citep[e.g.,][]{temim2013,eagle2022}. A spectral break in the X-ray band $\sim 10\,$keV would be expected.  Accurate characterization of the synchrotron cut-off in particular would not only improve the broadband interpretation for the PWN, but would provide insight to the MeV predicted pulsar and place new constraints on the photon fields contributing to the IC emission of the PWN.  

In order to constrain this spectral region further, deeper NuSTAR observations are required. Furthermore, it may be worth exploring the Fermi--LAT data using more data. The work of \citet{eagle2023} detects B0453--685 as a faint point-source with dominant emission detected between 1--10\,GeV using 11.5\, years of data. Today there is now $\sim 4$ more years of data, a 35\% increase in integration time in the Fermi band. Secondly, the MeV predicted pulsar may be investigated with future MeV space missions such as COSI\footnote{\url{https://cosi.ssl.berkeley.edu/}} (currently planned for launch in 2027)  and AMEGO-X\footnote{\url{https://asd.gsfc.nasa.gov/amego-x/}} (currently planned for launch by end of 2028). Finally, the Cherenkov Telescope Array\footnote{\url{https://www.cta-observatory.org/}} will be able to provide additional constraints on the IC emission spectra reported in \citet{eagle2023}. 

Finally, we report the detection and characterization of a high-mass X-ray binary candidate in both Chandra and NuSTAR datasets. A short optical monitoring campaign with the LCO to measure any variability of the source is reported and a periodic variability between $25-29$\,days is found, 
suggesting the source is a high mass X-ray binary system with an orbital period equal to the measured periodic variability. Deeper optical and X-ray observations may confirm the true nature of this source. 



\acknowledgements
This work has made use of data from the European Space Agency (ESA) mission
{\it Gaia} (\url{https://www.cosmos.esa.int/gaia}), processed by the {\it Gaia}
Data Processing and Analysis Consortium (DPAC,
\url{https://www.cosmos.esa.int/web/gaia/dpac/consortium}). Funding for the DPAC
has been provided by national institutions, in particular the institutions
participating in the {\it Gaia} Multilateral Agreement. J. H. acknowledges support from NASA under award number
80GSFC21M0002.  Work at NRL is supported by NASA.
\\
\\

\bibliographystyle{apj}
\bibliography{aa.bib}

\end{document}